
\magnification=1200

\def \im {\rm Im}

\def\text#1{\hbox{\rm #1 }}

\medskip
\line {\hfil DAMTP R93/26}
\line {\hfil gr-qc/9310016}

\vfill
\centerline {\bf SEMI-CLASSICAL LIMITS OF}
\smallskip
\centerline {\bf SIMPLICIAL QUANTUM GRAVITY}

\vfill

\centerline { \bf  J.W. BARRETT$ ^{(1)}$}
\medskip
{\it
\centerline { Department of Mathematics,}
\centerline { University of Nottingham,}
\centerline { University Park,}
\centerline { Nottingham, NG7 2RD, U.K.}}

\vfill

\centerline { \bf T.J. FOXON$ ^{(2)}$}
\medskip
{\it
\centerline { Department of Applied Mathematics and Theoretical Physics,}
\centerline { University of Cambridge,}
\centerline { Silver Street,}
\centerline { Cambridge, CB3 9EW, U.K.}}

\vfill

\noindent {\bf Abstract.}\smallskip

   We consider the simplicial state-sum model of Ponzano and Regge as a
path integral for quantum gravity in three dimensions.\smallskip

  We examine the Lorentzian geometry of a single 3-simplex and of a
simplicial manifold, and interpret an asymptotic formula for $6j$-symbols
in terms of this geometry. This extends Ponzano and Regge's similar
interpretation for Euclidean geometry.\smallskip

   We give a geometric interpretation of the stationary points of this
state-sum, by showing that, at these points, the simplicial manifold may be
mapped locally into flat Lorentzian or Euclidean space. This lends weight to
the
interpretation of the state-sum as a path integral, which has solutions
corresponding to both Lorentzian and Euclidean gravity in three dimensions.

\vfill

\vbox{ \parindent=0pt
e-mail address:

$(1),$ jwb@maths.nott.ac.uk,

$(2),$ tjf12@phx.cam.ac.uk .}

\eject

\beginsection { 1. Introduction.}

  The simplicial approach to gravity dates back to Regge [2], who
was the first to formulate a discrete form of
general relativity on a $n$-dimensional simplicial manifold.
He showed that $n$-dimensional curved space-time
could be built up from joining together flat $n$-dimensional simplicial blocks,
where the curvature is measured by the deficit angle $\varepsilon_l$ around
an $(n-2)$-dimensional bone (simplex) of area $A_l$. The Regge
action,
 $$ S_R = {1 \over 8\pi}\sum_{{\rm bones}, l} A_l \; \varepsilon_l,
\eqno(1.1) $$
is the discrete action which  approximates the Einstein-Hilbert
action of  general relativity.

   Ponzano and Regge [1] studied a model in which the simplicial blocks are
3-dimensional tetrahedra.
Each edge of a tetrahedron is labelled by a half-integer $j$, corresponding to
the $(2j+1)$-dimensional fundamental representation of the group $SU(2)$,
and the length of that edge is assigned to be $(j + {1\over 2}) \; \hbar $. A
tetrahedron is then a geometric representation (Figure 2) of the Racah
co-efficient, or $6j$-symbol, which is the recoupling co-efficient between
four angular momenta (see Figure 1 ).
Regge and Ponzano proposed expressions for different semi-classical limits
of a $6j$-symbol. These limits correspond to taking $\hbar \to 0$, while
$ j \to \infty$, such that each edge length $(j + {1 \over 2}) \; \hbar $
remains constant.

Regge and Ponzano then preceded to build up a general 3-dimensional
simplicial manifold out of tetrahedra. Their crucial insight was to consider
the state sum over labellings $j_l$ of the internal edges of the product
of $6j$ symbols, one for each tetrahedron $t$, weighted
by a specified function $f(j_l)$. This defines the partition function
$$ Z_{RP} = \sum_{{\rm labellings,} j_l} \Biggl( f(j_l) \prod_{{\rm tetra},t}
    \biggl\{ \matrix
    { j_1^{(t)} & j_2^{(t)} &j_3^{(t)} \cr j_4^{(t)} & j_5^{(t)} & j_6^{(t)}
\cr}
    \biggr \} \Biggr).
    \eqno (1.2)$$

 By taking a certain semi-classical limit of this partition function, they
argued that it has many of the features of a path integral for 3-dimensional
gravity, i.e.,
$$  \int d\mu \exp\; (iS_R),\eqno(1.3)$$
where $S_R$ is the Regge action $(1.1)$ for discrete Euclidean gravity in
$(3+0)$ dimensions.

  Stationary phase solutions of the semi-classical limit of the partition
function should then correspond to classical geometries. Regge and Ponzano
showed that some cases of stationary solutions correspond to flat Euclidean
3-space. Further details of this model and its relation to Penrose's spin
networks [3,4] were elucidated by Hasslacher and Perry [5].

  These ideas have been recently rediscovered in the context of
topological field theories. In particular, the theory of Turaev and Viro
[6] is the generalization [7,8] of the Regge-Ponzano
model, in which spins are replaced by representations of the quantum
group $U_q(sl(2))$. The semi-classical limit of the
sum-over-representations in the Turaev-Viro model has been shown [9]
to be related to a path-integral for 3-d Euclidean gravity with a
cosmological constant, which is related to the level of the Turaev-Viro
theory.

   In this paper we complete Regge and Ponzano's analysis of the stationary
points of the actions which arise in the semi-classical limit, giving each of
them a geometrical interpretation. Two types of stationary points arise,
having Euclidean geometry or
Lorentzian geometry. We analyse the case of Lorentzian geometry in detail, as
the basic facts are not well-known. The case of Euclidean geometry is similar.

In sections 2 and 3 we analyse the Lorentzian geometry of a simplex, and show
that Regge and Ponzano's exponential asymptotic formula has an interpretation
in terms of the Lorentzian geometry of a single simplex. This result is
entirely analogous to Regge and Ponzano's geometric interpretation of their
oscillatory asymptotic formula in terms of Euclidean geometry.

In sections 4 and 5 we consider the relation of some of the stationary
points of the state
sum model for a simplicial manifold to Lorentzian geometry.
Our main result is in section 5. This is that the stationary points of the
action, in the semiclassical limit, correspond to a generalised flat Lorentzian
space. In this generalisation, there is
locally a mapping of a neighbourhood of the simplicial manifold to flat
Minkowski space. This is a generalisation of the
usual notion of a flat Lorentzian metric on a manifold which determines a local
1-1 mapping of the simplicial manifold
to a region of Minkowski
space. The condition that it is locally 1-1 is dropped
for our generalised flat solutions. In the semiclassical limits which
correspond to Euclidean space, there is an entirely analogous analysis.

The physical interpretation of these results is discussed briefly in section
6.

\beginsection {2. Lorentzian geometry of a simplex.}

  In this section, we  consider the Lorentzian geometry of a tetrahedron. We
define the analogues of angles between faces of a tetrahedron in Lorentzian
space, for the particular cases which arise in the Regge-Ponzano model.
As described in the next section, a condition imposed in this model is that
the values on the edges around a  face satisfy the triangle inequalities. This
implies that the  tetrahedra which arise are those in which all faces, and
hence
all edges, are spacelike.

  First consider the usual definition of angles in Euclidean space.
Consider a vertex of a triangle. The interior angle is the
angle between the two sides, and the exterior angle is the angle between the
two normals. These are both numbers between 0 and $\pi$ and  it follows  from
the definitions that
$$ \text {interior angle} = \pi - \text {exterior angle}. \eqno(2.1) $$

  Now consider a vertex of a triangle in 2-dimensional Minkowski space
which has all its edges spacelike. Suppose the vertex is at the origin, for
convenience. There are two cases to consider
 \item {(1)} Thin wedge. The triangle does not contain any of the light cone of
the vertex (See Figure 3, which depicts the triangle, light cone and the
normals to two of the edges). There is an element of $SO(1,1)^+$ which takes
one edge
into another. The Lorentzian angle at the vertex is defined to be the boost
parameter of this element, which is a positive real number. This is an interior
angle; an exterior angle is not defined.

\item {(2)} Thick wedge. The triangle contains all of the future, or all of
the past portion of the light cone (Figure 4). In this case there is an element
of $SO(1,1)^+$ which takes one outward normal into the other outward
normal. The Lorentzian angle at the vertex is defined to be minus the boost
parameter of this element, and so is a negative real number. This is an
exterior angle; an interior angle is not defined.
\medskip

Thick and thin wedge angles are related by the following
construction. Starting with a thick wedge (say), produce one edge a distance
past the vertex. The line segment so added forms two sides of a thin wedge
triangle with the other edge of the original triangle (Figure 4). It is
straightforward to see that the thick and the thin wedges have opposite angles,
i.e., one is minus the other.

Now consider a tetrahedron in 3-dimensional Minkowski space, which has all
edges and all faces spacelike. Taking an orthogonal cross-section through an
edge yields one of the wedges discussed above. The Lorentzian angle between
the two faces which meet at the edge is defined to be the angle of this wedge.
It follows that these angles are also one of two types, interior or exterior.

The Lorentzian angle between two faces of a tetrahedron meeting at an edge
$(hk)$ can be expressed in terms of the normals $n_h$ and $n_k$ to the
faces (using the convention for the signature of the metric that $n$ is
timelike
with norm $n^2 = -1$). For an interior angle (Figure 3), it is given by
$$ \Theta_{hk} = \cosh^{-1} \; ( n_h . n_k), \eqno(2.2)$$
and for an exterior angle, (Figure 4), by
$$ \Theta_{hk} = - \cosh^{-1} \; (-n_h . n_k). \eqno(2.3)$$

 Only certain patterns of interior and extrerior angles can occur over a given
simplex. Since the faces are spacelike, their normals are timelike and hence
can
be classified as future-pointing or past-pointing. Two different types of
Lorentzian simplex can now be distinguished. One type has two future-pointing
normals to faces, and two past-pointing. The other type has three
future-pointing and one past-pointing, or vice versa. The angle between two
faces with normals both pointing in the same sense is exterior, whereas if one
normal is future-pointing and one past pointing, the angle is interior. Thus,
the
first type of Lorentzian simplex has three exterior and three interior angles,
and the second type has two exterior and four interior angles.

\beginsection {3. Asymptotic formulae for a 3-simplex.}

  Regge and Ponzano [1] proposed two asymptotic formulae for $6j$-symbols.
They interpreted  one of these in terms of
the geometry of a simplex in Euclidean space. In this section, we show how the
other case may be interpreted in terms of the Lorentzian
geometry discussed in the previous section.

 Regge and Ponzano associated a tetrahedron (3-simplex) to a
$6j$-symbol, by defining the edge-lengths to be
$j_{12} = (a + {1\over 2}) \; \hbar, \;  j_{13} = (b + {1\over 2}) \; \hbar, $
etc. (Figure 2), corresponding to the $6j$-symbol
$$ \biggl \{ \matrix { a & b & c \cr d & e & f \cr} \biggr \} .$$

The $6j$-symbol is only defined when the values of the angular momenta
which correspond to the edges around a face satisfy the triangle inequalities,
$ | b - c | \leq a \leq b+c $. This implies that the edge-lengths around a face
satisfy
the triangle inequalities, up to additional terms of $\pm {1 \over 2}$.
Also, $a,b,c$ are required to satisfy $a+b+c=\hbox{integer}$, for each face.
If these constraints are not satisfied, the value of the $6j$-symbol is
defined to be zero.

  The vertices and edges of a tetrahedron define a graph. The six edge lengths
$j_{hk}$, for $h<k=1,\ldots 4$, determine distances between the vertices of
the graph. The geometric problem is to find an embedding of this graph into
flat Euclidean or Minkowski space.

  The volume $V$ of the tetrahedron is given by the Cayley determinant

$$ V^2 = {1 \over {2^3 \;  (3!)^2} }
           \left | \matrix { 0 & {j_{34}}^2 & {j_{24}}^2 & {j_{23}}^2 & 1 \cr
                                     {j_{34}}^2 & 0 & {j_{14}}^2 & {j_{13}}^2 &
1 \cr
                                     {j_{24}}^2 & {j_{14}}^2 & 0 & {j_{12}}^2 &
1 \cr
                                     {j_{23}}^2 & {j_{13}}^2 & {j_{12}}^2 & 0 &
1 \cr
                                      1 & 1 & 1 &1 & 0 \cr }
            \right | \; .
            \eqno (3.1)
$$

     The tetrahedon may be embedded in a 3-dimensional Euclidean space if
and only if $ V^2 > 0 $.
However,  the triangle inequalities around each face are not sufficient to
imply this. In fact, it is also possible to have $ V^2 < 0 $, in which case the
tetrahedron may be embedded in 3-dimensional Minkowski space.

  Note that the volume $V$ is related to the length $j_{hk}$, the normals
$n_h$ and $n_k$  and the areas  $A_h$ and $A_k$ of the faces meeting at this
edge, by the formula

$$  V^2 =  { 4 {A_h}^2 {A_k}^2 \over 9 {j_{hk}}^2 } ( 1 - (n_h . n_k)^2 ),
    \eqno(3.2) $$
implying that $V^2 \le 0$, if any of, and hence all of, the pairs of normals
satisfy $|n_h . n_k|>1$.

  Regge and Ponzano proposed expressions for the asymptotic limit
of a $6j$ symbol in which all the angular momenta become large, for each of
these two cases. If the angular momenta are scaled by a factor $\lambda$,
this limit corresponds to taking $\lambda \to \infty$, while $\hbar \to 0$,
such that the edge lengths $(\lambda a + {1 \over 2}) \; \hbar $
remain constant. They argued that the asymptotic limit for $V^2 > 0$ is

$$ \biggl \{ \matrix { a & b & c \cr d & e & f \cr} \biggr \} \simeq
     {1 \over \sqrt{ 12 \pi \; V}} \;
   \cos \biggl( \sum_{h < k} j_{hk} \; \theta_{hk} + {\pi \over 4} \biggr),
   \eqno (3.3)
$$
where $\theta_{hk}$ is the exterior angle between the outward normals of the
faces meeting at edge $(hk)$, $j_{hk}$ is the length of this edge, and the sum
is
over the six edges of the tetrahedron, labelled by $(hk)$, with $h<k$.

   By setting up a formal analogy with the WKB method, Regge and Ponzano
were able to continue their asymptotic formula to the region $V^2 < 0$, to
obtain

$$  \biggl \{ \matrix { a & b & c \cr d & e & f \cr} \biggr \} \simeq
     {1 \over 2 \sqrt{ 12 \pi \; |V|  }} \;
     \cos \phi \;
     \exp \biggl( - \biggl | \sum_{h < k} j_{hk} \; { \im} \; \theta_{hk}
     \biggr | \biggr) ,
     \eqno (3.4)
$$
where $\theta_{hk}$ is still defined by the Euclidean formula
$$ \cos \theta_{hk} = {n_h . n_k\over n_h^2}=-n_h . n_k. \eqno(3.5)$$
This implies that $\theta_{hk}$ has the form
$$ \eqalignno
     {\theta_{hk} &= m_{hk}  \pi + i \; {\rm Im} \; \theta_{hk}, & (3.6) \cr}
$$
for an integer $m_{hk}$, which is fixed by putting
$$ \eqalignno {m_{hk} &= \cases { 0, &{exterior angle about $(hk)$,} \cr
                                  1, &{interior angle about $(hk)$.} \cr}
                 &(3.7) \cr}$$
The number $\phi$ is defined by
$$\eqalignno{ \phi  &=  \sum_{h < k} \; ( j_{hk} - {1\over 2}) \;  m_{hk}
\pi.  & (3.8) \cr }$$
These definitions imply that for allowed values of $j_{hk}$,
i.e. where the
$6j$-symbols are non-zero,
$$ \cos \phi = (-1)^{ \bigl(\sum (j_{hk} -{1\over 2}) \; m_{hk} \bigr)} \; .
     \eqno(3.9)$$

   We shall now show that the formula (3.4) for the asymptotic limit has a
natural interpretation in terms  of the embedding of a tetrahedron with
$V^2 < 0$ in Lorentzian space, in  which the imaginary parts of angles are
interpreted in terms of the Lorentzian angles, defined in section 2.

 From the definitions of Lorentzian angles $\Theta_{hk}$ in equations (2.2) and
(2.3),
$$  n_h . n_k = -(-1)^{m_{hk}} \; \cosh \Theta_{hk} \; . \eqno(3.10) $$
However, we want to identify the Lorentzian angle with the
imaginary part of the Euclidean angle $(3.6)$.

 From equation (3.6),
$$ \eqalignno {
     \cos \theta_{hk} &= \cos \; ( m_{hk} \pi + i \; {\rm Im} \; \theta_{hk})
\cr
     &= (-1)^{m_{hk}} \; \cosh \; ({\rm Im} \; \theta_{hk}).
     & ( 3.11) \cr } $$
Thus, we may put
$$ \Theta_{hk} = {\rm Im} \; \theta_{hk} \; . \eqno(3.12) $$

   The asymptotic limit for the $6j$-symbol, in which all the angular
momentum become large, for the case $V^2 < 0$ is therefore equal to

$$  \biggl \{ \matrix { a & b & c \cr d & e & f \cr} \biggr \} \simeq
     {1 \over 2 \sqrt{ 12 \pi \; |V|  }} \;
     \cos \phi \;
     \exp \biggl( - \biggl |  \sum_{h < k} j_{hk} \;  \Theta_{hk}
     \biggr | \biggr) . \eqno (3.13)
$$

   This formula for the asymptotic limit in the Lorentzian case is now in the
same form as that for the limit in the Euclidean case (3.3). In the Euclidean
case, Regge and Ponzano interpreted $\sum j\, \theta$ as the Regge action for
the simplex. It seems reasonable for us to interpret $\sum j \, \Theta$ as the
Regge action for a Lorentzian simplex, as we shall show in the next section.

\beginsection {4. Lorentzian geometry of a simplicial manifold.}

  We now want to consider the Lorentzian geometry of a  3-dimensional
simplicial manifold, consisting of 3-simplices of the type which
may be embedded in Lorentzian space. The curvature of the simplicial manifold
is measured by the deficit angles around 1-dimensional edges in the complex,
in terms of which the Regge  action is defined.

   Firstly, recall that for a  $n$-dimensional Euclidean simplicial manifold,
Regge defined the deficit angle $\varepsilon_l $, around an
$(n-2)$-dimensional
`bone' of area $A_l$, (see Figure 5), by
 $$  \varepsilon_l = 2\pi - \sum_{{\rm tetra,}t} \alpha_l^t, \eqno(4.1) $$
where $\alpha_l^t$ are the interior angles in the simplices around the
bone.
 For flat space, the sum of the interior angles around
the bone will be $2\pi$, implying that the deficit angle is zero.

   For the case of a Lorentzian simplicial manifold, we define
the Lorentzian deficit angle $\varepsilon_l $ to be (see Figure 6)
$$  \varepsilon_l = \sum_{{\rm tetra,}t}  \; \Theta_l^t, \eqno(4.2) $$
where $\Theta_l^t$ are the Lorentzian angles in the simplices around the
bone, which we defined in section 1.
With this definition, the deficit angle for flat
Lorentzian space is also zero.

   The Regge action is then defined for either case by
$$ S_R = \kappa \sum_{{\rm edges},l} A_l \; \varepsilon_l. \eqno(4.3) $$
This is the discrete action given by evaluating the Einstein-Hilbert
action on the simplicial manifold. The constant $\kappa$ is
${1 \over 8\pi}$ for general relativity. We shall take $\kappa=1$
for convenience.

   For the 3-dimensional case, the bones are the 1-dimensional edges of
length $j_l$, so that
the 3-dimensional Regge action for a Lorentzian simplicial manifold is
$$ \eqalignno{ S_R &=   \sum_{{\rm edges},l} j_l \;
             \varepsilon_l,  \cr
             &=  \sum_{\rm tetra,t}\biggl(\sum_{\rm
edges,l}j_l\Theta_l^t\biggr).
             &(4.4) \cr
}$$
This means that $\sum j\Theta$ may be
interpreted as the Regge action for a  single tetrahedron.

\beginsection {5. 3-dimensional simplicial gravity.}

   As explained in the introduction, our motivation for considering Lorentzian
simplicial geometry is to extend the work of Regge and Ponzano [1] to the case
of 3-dimensional Lorentzian simplicial gravity. Regge and Ponzano showed that
the semi-classical limit of the state-sum over representations of products of
$6j$-symbols, corresponding to tetrahedra for which $V^2 > 0$, may be
interpreted as a path-integral for 3-dimensional Euclidean gravity. This
state-sum is thought of as a discrete version of the sum over metrics in the
standard approach to path-integral quantum gravity [15].

   We have shown that the semi-classical limit of a single $6j$-symbol,
corresponding to a tetrahedron for which $V^2 < 0$, has a natural
interpretation
in terms of embedding that 3-simplex in a 3-dimensional Lorentzian space. A
complex of such 3-simplexes forms a simplicial manifold, on which the Regge
action for gravity is defined. We now want to ask if the semi-classical limit
of
the state-sum of products of the corresponding $6j$-symbols may be interpreted
as a path-integral for 3-dimensional Lorentzian gravity with this action.

   The Regge-Ponzano model consists of a simplicial manifold formed by  a
complex of 3-simplexes.  A labelling of this simplicial manifold is defined by
assigning a representation $a$ of $SU(2)$ to each internal and external edge of
the complex, such that the $6j$-symbol associated to each 3-simplex
is well-defined, by virtue of satisfying the necessary conditions described in
section 2.  For a given labelling, the length of each edge is taken to be
$j = (a+ {1\over 2}) \hbar$.  The initial data for the model consists of a
particular labelling of all edges in the boundary. To relate this to gravity,
recall [12] that specifying all external edge-lengths is equivalent to
specifying a 2-metric on the boundary.

   For a given initial labelling of the boundary, there are an (infinite)
number of
possible ways of consistently labelling the internal edges of the complex. A
state of the model is one such labelling of the $n$ internal edges, denoted by
$\{x_l \}, \; l= 1, \ldots, n$.
  The Regge-Ponzano partition function is defined as the sum, over all possible
states, of products of $6j$-symbols $T_t=\left\{\matrix{. . .\cr. . .}\right
\}$, which correspond to each tetrahedron
$t$. It is given by
$$ Z_{RP} = \sum_{{\rm state,} \{x_l \} } \; A(x_1, \ldots, x_n),\eqno (5.1)$$
where
$$ \eqalignno{ A(x_1, \ldots, x_n) &=  \prod_{{\rm int. edges},l}
     (2 x_l + 1) \; (-1)^{\chi}
     \prod_{{\rm tetra},t}  T_t,
    &(5.2) \cr \cr
   \chi &= \sum_{{\rm edges,}l} (n_l - 2) \; x_l, &(5.3) \cr
}$$
and  $ n_l$ is the  number of simplexes meeting at edge $ l$.

   The key point is that this partition function is independent of the way that
the interior of the manifold is triangulated, and depends only on the initial
data on
the boundary and the topology of the interior. Hence, it formally defines a
topological field theory.

   However, if there are any internal edges, then the partition function is
infinite and must be regularised to give finite answers. A simple
regularisation procedure was demonstrated by Regge and Ponzano [1]. The
partition function is finite for cases where the manifold has a boundary,
the vertices are all in the boundary and any loop in the manifold can be
deformed to a loop on the boundary. This is because the triangle inequalities
constrain the labellings of the interior edges to a finite set of values.
Examples of such manifolds are the handlebodies, which are the solid sphere,
the solid torus, and solid surfaces of higher genus. It is remarkable that
Witten's discussion [13]  of quantised gravity, from a different viewpoint,
also arrived at the conclusion that the partition function is well-defined for
these manifolds but requires regularisation to make sense in the more general
cases.

  We shall not consider the regularisation in what follows. A more rigorous
approach would be to consider the Turaev-Viro model, in which the state
sum is always finite, as mentioned in the introduction.

  Our basic assumption is that the partition function (5.1) is dominated by
stationary phase points or saddle points, for which a WKB-type approximation
is valid. Regge and Ponzano's asymptotic formulae for $6j$-symbols, (3.3),
(3.13), give two distinct regions in which we can reasonably look for such
stationary points.  The first such region is where all $j$'s are large and
every simplex is Euclidean, as considered by Regge and Ponzano. The second
region is where all $j$'s are large and every simplex is Lorentzian. We
consider the Lorentzian region first.

    Taking the Lorentzian semi-classical limit $(3.13)$ of each $6j$-symbol,
the partition function  becomes a sum over terms of the form

$$  A(x_1, \ldots, x_n) =  \prod_{{\rm int. edges},l} (2 x_l + 1) \;
     (-1)^{\chi}
     \prod_{{\rm tetra},t} {\cos \phi_t \over 2 \sqrt{ 12 \pi \; |V_t|  }}\;
     \exp \biggl( - \Bigl |  \sum_{{\rm edges,} k} j_{k} \;  \Theta_{k}^t
     \Bigr | \biggr). \eqno (5.4)$$

  The important feature is the behaviour of $A(x_1, \ldots, x_n)$ as each
$x_l$ varies. Because the sum of the angular momenta around a face is
an integer, the $x_l$ may only vary by integers.
Under  $x_l\mapsto x_l + 1$, the term
$$ (-1)^{\chi} \prod_t \cos \phi_t, \eqno (5.5)$$
varies by multiplication by  $(-1)^{N_{k}}$,
where $N_{k}$ is the number of thick wedges around the $k$-th edge.
Stationary points are therefore only possible when $N_{k}$ is even.
The remaining terms are all positive, so the sum will be dominated by
 points where the exponent in
$$
     \exp \biggl( - \sum_{{\rm tetra,}t} \; \Bigl |  \sum_{{\rm edges,} k}
j_{k} \;
     \Theta_{k}^t
     \Bigr | \biggr) \eqno(5.6)$$
is greatest and is slowly varying. In such a situation a number of
neighbouring terms are all large. This is the discrete version of
the steepest descent approximation to a contour integral. The exponent in (5.6)
defines an action $S_Q$, which has the same form as the Regge action (4.4),
except
that, here, the modulus of the action for each tetrahedron is  taken.

The first question is whether this modulus sign is really necessary. We
have checked that it is. Numerical experiments show that action
$\sum j\Theta$ is negative for the tetrahedra with two future pointing
faces, and positive for tetrahedra with one or three future pointing
faces. However, it is not known whether this is true in general.
Also, we checked that the Regge-Ponzano formula is a good approximation
in both cases; surprisingly, all the numerical tables in their paper have
only tetrahedra of one type. This is because it is quite hard to find
small half-integers for which the tetrahedron has three or one future pointing
normals.

The next question is to examine the stationary points of the action $S_Q$
and discuss their geometric significance. For each tetrahedron $t$, define
$$  \sigma_t = {\rm sign} \Bigl (  \sum_{{\rm edges,} k} j_{k} \; \Theta_{k}^t
     \Bigr ). \eqno(5.7)
$$
According to our numerical experiments, these signs do not change
as the $j$'s are varied within one type of Lorentzian simplex.
The variation of the action
$$ S_Q = \sum_{{\rm tetra,}t} \; \sigma_t \; \Bigl ( \sum_{{\rm edges,} k}
     j_{k} \; \Theta_{k}^t \Bigr ) \eqno(5.8)$$
with respect to the length of edge $l$ is therefore
$$ {\delta S_Q \over \delta j_l} = \sum_{{\rm tetra,}t} \; \sigma_t \;
    \Theta_{l}^t \; + \sum_{{\rm tetra,}t} \; \sigma_t \Bigl
    (\sum_{{\rm edges,} k}
    j_{k} \; {\partial \Theta_{k}^t \over \partial j_l} \Bigr),
    \eqno (5.9) $$
The second term is zero by the Lorentzian version of Regge's calculation [2].
Hence, the stationary points of the action are given by one equation for each
edge $l$
$$ \sum_{{\rm tetra,}t} \; \sigma_t \;
    \Theta_{l}^t = 0. \eqno (5.10) $$

  We claim that this is the required condition for the simplicial manifold to
be
mapped locally into  a flat Minkowski space.
This shows that the stationary configurations  are locally flat in a
generalised sense. The normal definition of a flat manifold is that a
neighbourhood of each point is mapped by a 1-1 mapping to a region of
flat space. Here, the generalisation is that the mapping is not required to
be 1-1. The singular points of this mapping, i.e., the folds, lie on the
2-simplexes of the simplicial manifold. Each individual simplex is mapped
isometrically, but the singular nature of the map is manifested in that
several points, each lying in different simplexes, may be mapped to the same
point of Minkowski space.

   The argument that this condition is true is as follows. Consider the
neighbourhood of
an edge in the simplicial manifold. This consists of a number of
wedges from the tetrahedra that meet at that edge, arranged in a circle
about the edge. From (5.7), there is a sign $ \sigma_t =\pm 1$ associated to
each tetrahedron, and a Lorentzian angle $\Theta_l^t$ in each wedge at the
edge.
The neighbourhood is mapped to flat $(2+1)$-dimensional Lorentzian space by the
following  rule (see Figure 7). Pick one wedge and map it isometrically into
flat
space. The wedges are now treated in the order in which they circulate the
edge.
The  second wedge to be mapped into flat space is adjacent to the first wedge
in
the manifold, and it shares a common face. There are two possible ways of
mapping  this into the flat space so that the face is still common. In one of
these
ways, the wedges meet only on the common face. This is chosen if the signs
$\sigma_t$ are equal for the two tetrahedra. In the other way of mapping the
second wedge to flat space the two wedges overlap, and this is chosen if the
signs differ. In a sense, the second wedge backtracks the first. This process
is then repeated for the third and succesive wedges in a similar way.

Consider the two-dimensional normal cross-section to the edge, as in Figure 7.
The Lorentz transformation which takes the outer edge of the first wedge to the
outer edge of the last wedge is
$$ (-1)^{N_l}\pmatrix{\cosh\varepsilon'_l&\sinh\varepsilon'_l\cr
\sinh\varepsilon'_l&\cosh\varepsilon'_l}\eqno(5.11)$$
(or its inverse), where
$$ \varepsilon'_l = \sum_{{\rm tetra,}t} \; \sigma_t \;
    \Theta_{l}^t.  \eqno(5.12)$$

Therefore, when all of the wedges have been mapped into flat space the first
and the last wedges meet on their common face to form a continuous whole if and
only if this matrix is the identity matrix. This condition is equivalent to
the vanishing of  $\varepsilon'_l$ and the requirement that $N_l$ be even.
These
are the two conditions for stationary points in the state sum model.

We may regard $\varepsilon'_l$ as a generalised deficit angle and
(5.10) as the condition for this generalised deficit parameter to be zero.
  Our conclusion is that the stationary points of the partition function (5.1),
in the
region of large $j$ on Lorentzian tetrahedra, correspond to a Lorentzian space
which is locally flat in the generalised sense. This means that there is a map
from
the manifold to flat Minkowski space.

In the Euclidean region for stationary points, the approximation is obtained
by putting formula (3.3) in the state sum (5.1). This is discussed by
Moussouris
[10] and Ooguri [11]. Each cosine in (3.3) may be written as a sum
$$\cos \alpha =
e^{i\alpha} + e^{(-i\alpha)}. \eqno(5.13)$$
This leads to stationary phase points, as
opposed to maxima which were encountered for the Lorentzian case. However, the
same interpretation may be given. Each simplex may be assigned a sign, $
\sigma_t
= \pm 1$, and the geometry of a stationary phase point is a generalised flat
Euclidean space. There are terms for each possible assignment of signs to each
tetrahedron, and each of these may give rise to a stationary point. The
stationary
points do not just correspond to taking all of the terms with the same sign,
$e^{i\alpha}$, for example.

\beginsection{6. Discussion.}

  Our interpretation of the stationary points for the Regge-Ponzano partition
function is that they correspond to generalised flat, three-dimensional
Lorenzian or Euclidean spaces. This lends weight to the idea that this
partition function may be considered as a discrete version of the path
integral for quantum gravity in three dimensions.  Obviously, this idea is by
no means complete. In particular, the state sum is infinite for the
Regge-Ponzano model. It would be interesting to see if our interpretation goes
through to the Turaev-Viro model [6], in which the partition function is given
by a state sum over representations of a quantum group at a root of unity, and
so is naturally finite. For that model, we would expect the stationary points
to correspond to generalised Lorentzian or Euclidean spaces of constant
curvature in three dimensions. However, it is encouraging that a discrete
model can be found with a partition function which has stationary solutions
corresponding to flat space, as we would expect a path integral for
3-dimensional quantum gravity to be concentrated on these solutions. Penrose
has
argued forcefully  [3,4] that the best approach to quantum gravity is to look
for a discrete  model from which space-time emerges in some limit.

  The generalised flat solutions for the Einstein equations were discussed by
Horowitz [14]. They arise naturally as the solutions to the Einstein
equations when the Einstein action is written in terms of the frame field and
connection form. In three dimensions, such a stationary point is precisely
a flat $ISO(2,1)$ connection and a section of the associated bundle with
fibre affine Minkowski space.
This data is equivalent to a map from the space-time manifold to a flat
Lorentzian space. As in this paper, the mapping need not be 1-1.
The target manifold need not have the same topology, and of course
the mapping will induce a degenerate metric on the original space-time.
Ooguri [11]  suggested that the Regge-Ponzano theory should be interpreted
directly in terms of the path integral for the action in this form, i.e., with
frame fields and connection forms. However,
our interpretation and Horowitz' results only imply such an interpretation for
the stationary points.

It is interesting to consider whether the stationary points considered in this
paper have an interpretation as a map to another manifold which has a flat
Lorentzian metric, in other words whether our local characterisation
extends to this global characterisation.

   In the standard approach to quantum gravity using path integrals [15], the
path
integral is the sum over metrics on the interior of a manifold, consistent with
the
metric on the boundary. However, in this case, the path integral is not
well-defined
since the Einstein action is unbounded below for conformal rescalings. This
problem may
be solved by ``the rotation of the contour for the conformal factor" [16]. This
is the
formal (i.e. informal) process of redefining the path integral for the
conformal
fluctuations separately, which has the effect of changing the sign of the
fluctuation
of the action for these modes. This process renders the action bounded below.
However, this is precisely what happens
at our Lorentzian semiclassical approximation: the action is strictly positive,
because of the modulus in equation (5.6).
The price of maintaining a geometric description of the action is that the
classical
solutions have the feature that there are regions in which
``time goes backwards''. Indeed, since each face of each tetrahedron is
spacelike,
if two neighbouring tetrahedra have opposite signs $\sigma_t$, then
they have opposite induced time orientations from the Minkowski space.

The state sum model can be regarded as a definition of a path integral for
quantum gravity.
 The terms in the state sum (5.1) which are not stationary do not correspond
to classical geometries. This is a typical feature of path integrals.
However, path integrals have approximations around stationary points which
have a geometric interpretation.
The fluctuations in the state-sum model around a Lorentzian stationary point
may be written, according to the asymptotic approximation, as
$$ \eqalignno {
     Z_{RP} &= \int \prod_l {\rm d} \mu (j_l) \; \exp (- S_Q),   &(6.1)
\cr} $$
where the generalised Regge action is
$$S_Q=\sum_{{\rm edges},k} j_k \; \varepsilon_k' \; .  \eqno(6.2) $$
This looks like a path integral in the sense of integration of a space of
metrics with an action.
Note, however, that contrary to what we would
normally expect for a Lorentzian path-integral, the integral is over terms of
the form $e^{-S}$ rather than $e^{iS}$. Also, in the
Euclidean case the corresponding result is an integral over $e^{iS}$ terms.

  As a final thought, this makes the state-sum model rather similar to quantum
cosmology models [17], in that the one path integral contains both Lorentzian
and Euclidean regimes. However, here, the Euclidean
regime is oscillatory and the Lorentzian regime exponential.

\beginsection{ \bf Acknowledgements.}

   TJF would like to thank Ruth Williams and Chris Fewster for helpful
discussions. TJF is supported by an S.E.R.C. research studentship.

\beginsection {\bf References.}

\noindent [1] G. Ponzano and T. Regge, ``Semiclassical
limit of Racah coefficients",

in {\it Spectroscopic and group theoretical methods in physics,}
ed. F. Bloch,

(North Holland, Amsterdam, 1968).
\medskip

\noindent [2] T. Regge, ``General relativity without
co-ordinates,"

Nuovo Cimento {\bf 19} (1961), 558.
\medskip

\noindent [3] R. Penrose, ``Angular momentum: An approach to
combinatorial space-time",

in {\it Quantum theory and beyond,} ed. T.Bastin, (C.U.P., Cambridge,
1971).
\medskip

\noindent [4] R. Penrose, ``On the nature of quantum
geometry", in {\it Magic without magic,}

ed. J.R. Klauder, (Freeman, San Fransisco, 1972).
\medskip

\noindent [5] B. Hasslacher and M.J. Perry, ``Spin
networks are simplicial quantum gravity,"

Phys. Lett. {\bf 103B} (1981),  21.
\medskip

\noindent [6] V.G. Turaev and O.Y. Viro, `` State sum
invariants of 3-manifolds and quantum

$6j$ symbols," Topology {\bf 31} (1992), 865.
\medskip

\noindent [7] H. Ooguri and N. Sasakura, ``Discrete and
continuum approaches to three-dimensional

quantum gravity," Mod. Phys. Lett. {\bf A6} (1991), 3591.
\medskip

\noindent [8] F.J. Archer and R.M. Williams, ``The Turaev-Viro state-sum
model and 3-dimensional

quantum gravity," Phys. Lett. {\bf B273} (1991), 438.
\medskip

\noindent [9] S. Mizoguchi and T. Tada, ``3-dimensional gravity from the
Turaev-Viro invariant,"

Phys. Rev. Lett. {\bf 68} (1992), 1795.
\medskip

\noindent [10] J.P. Moussouris, ``Quantum models of space-time based on
recoupling theory",

D.Phil. Thesis, (Oxford University, 1983).
\medskip

\noindent [11] H. Ooguri, ``Partition functions and topology-changing
amplitudes in the 3D lattice

gravity of Ponzano and Regge," Nucl. Phys. {\bf B382} (1992), 276.
\medskip

\noindent [12] R. Sorkin, ``Time-evolution problem in Regge calculus,"

Phys. Rev. {\bf D 12} (1975), 385.
\medskip

\noindent [13] E. Witten, ``Topology-changing amplitudes in 2+1
dimensional gravity,"

Nucl. Phys. {\bf B323} (1989), 113.
\medskip

\noindent [14] G.T. Horowitz, ``Topology change in classical and quantum
gravity,"

Class. Quantum Grav. {\bf 8} (1991), 587.
\medskip

\noindent [15] S.W. Hawking, ``The path-integral approach to quantum
gravity," in

{\it General Relativity: an Einstein centenary survey,}
ed. S.W. Hawking and W. Israel,

(C.U.P., Cambridge, 1979).
\medskip

\noindent [16] G.W. Gibbons, S.W. Hawking and M.J. Perry,
``Path integrals and the indefiniteness

of the gravitational action,"
Nucl. Phys. {\bf B138} (1978), 141.
\medskip

\noindent [17] J.B. Hartle and S.W. Hawking, ``Wave function of the universe,"

Phys.Rev. {\bf D28} (1983), 2960.
\medskip

\bye